\documentclass[conference,a4paper]{IEEEtran}
\usepackage{float}
\usepackage{amsmath}
\usepackage{amsthm}
\usepackage{amssymb}	
\usepackage{graphicx}
\usepackage{subfigure}
\usepackage{enumitem}
\usepackage{algorithm}
\usepackage{stfloats}
\usepackage{cite}
\usepackage{cuted}
\usepackage{color}
\newcommand*{\J}{\jmath}%

\usepackage{amsmath,amssymb,mathtools,bm,etoolbox}
\usepackage[center]{caption}

\DeclarePairedDelimiterXPP\Aver[1]{\mathbb{E}}{[}{]}{}{
	
	#1
}

\newtheorem{my_theorem}{Theorem}

\newtheorem{my_lemma}{Lemma}

\title{RIS-THz Wireless Communication with Random Phase Noise and Misaligned Transceiver}

\author{
	\IEEEauthorblockN{ Omkar R. Durgada, Vinay Kumar Chapala, and  S.~M.~Zafaruddin}\\
	\IEEEauthorblockA{ Department of Electrical and Electronics Engineering, 
		BITS Pilani, Pilani Campus, Pilani-333031, Rajasthan, India
		\\ Email: \{f20200460, p20200110, syed.zafaruddin\}@pilani.bits-pilani.ac.in}
		\thanks{This work was supported in part by the Science and
		Engineering Research Board (SERB), Department of Science and Technology
		(DST), Government of India, under MATRICS Grant MTR/2021/000890 and Start-up Research Grant SRG/2019/002345.}
}

\begin{document}
	\maketitle 
	\begin{abstract}
	Existing research works on reconfigurable intelligent surfaces (RIS) based terahertz (THz) system ignores the effect of phase noise and employ the zero-boresight pointing errors model of the free-space optics channel in performance analysis. In this paper, we analyze the performance of RIS-THz transmission under the combined effect of channel fading, THz pointing error (TPE), and statistical phase noise due to imperfect phase compensation at each RIS element. First, we derive statistical results of the double $\alpha$-$\mu$ fading combined with the TPE and phase noise at individual RIS elements using single-variate Fox's function. Next, we use the multi-variate Fox's H-function representation to develop exact analytical expressions for the density and distribution functions of the resultant signal-to-noise ratio (SNR) of the RIS-THz link considering the accumulating propagation effect from all RIS elements. Using the derived statistical results, we analyze the exact and asymptotic expressions for the considered system's outage probability and average bit-error rate (BER). The analytical results show that the diversity order of the system is independent of phase noise, depends on the channel fading parameters $\alpha$ and $\mu$, and depends on the $\beta$ parameter of the TPE.

	\end{abstract}				
	\section{Introduction}	
The ongoing development of sixth-generation communications (6G) aims to deliver the next wave of wireless intelligence by serving stringent requirements such as high data rate, high reliability, low transmission latency, coverage, and security \cite{Yishi_2021}. Much attention has been drawn towards reconfigurable intelligent surfaces (RIS), which has become a spearheading candidate technology for 6G, as it brings together scientific novelty and coherence. RIS is an artificially designed planar metamaterial with numerous passive reflective elements designed to enable programmable manipulation of the phase, amplitude, frequency, and even polarization of incoming electromagnetic waves without the need for complex decoding and encoding operations \cite{Basar2019_access}. The RIS can promise to overcome high attenuation by creating a virtual line-of-sight path cost-effectively through reduced hardware costs and energy consumption, specifically  for higher frequency mmWave and THz communications \cite{Basar2019_access,Qingqing2021}. 
	
Channel fading including the  higher atmospheric loss, misalignment errors, and phase noise are major bottlenecks for the performance of RIS-assisted THz transmission. An accurate performance assessment requires tractable and experimentally verified statistical models for  these channel impairments.   Recent  experiment campaigns validate  $\alpha$-$\mu$ distribution for the short term fading in THz link at a $142$ \mbox{GHz} carrier frequency for a shorter link \cite{scientific_report21}.  In addition to the channel fading, there occurs a misalignment error known commonly as pointing errors due to the highly directional nature of THz communication links and the dynamic positional behavior of transceiver antennas \cite{Sarieddeen_2021}. Due to the lack of statistical model for pointing errors specific to THz system,  initial research on RIS-THz \cite{du2020_thz, chapala2021THz} assumed zero-boresight pointing errors of free-space optics (FSO) \cite{Farid2007}, which caters to only a special case of THz transmission. It should be mentioned the mathematical representation of the zero-boresight  pointing errors facilitates a straightforward analysis when combined with the $\alpha$-$\mu$ fading \cite{Boulogeorgos_Error,Pranay2021, Pranay_2021_TVT,Li_2022_Pointing_Error,Badarneh2022_Pointing_Error}.  It is certain that pointing errors in a RIS-aided THz system is not trivial by any means motivating the authors  in \cite{Dabiri_Pointing_Error_2022} to develop  a new generalized pointing error model for mmWave and THz frequency directional antennas.  The proposed model is heedful of several hindrances and provides a general case of direct communication between two asymmetric and unstable transceiver. However, the new statistical model of pointing errors for THz link termed as THz pointing errors (TPE) is more involved mathematically restricting the authors  in \cite{Dabiri_Pointing_Error_2022} to present approximate analysis when combined with the $\alpha$-$\mu$ fading model. 

	In the few papers that adhere to RIS-aided THz systems  \cite{du2020_thz, chapala2021THz}, the authors have considered the combining effect of fading and zero-boresight pointing errors and ignored the effect of phase noise. It is necessary to understand that phase noise plays a pivotal role in the performance analysis pertaining to RIS-aided wireless systems and is a necessity for research outcomes to have relevance to practical applications. The aforementioned research brings out results on the assumption of ideal phase compensation at the RIS. However, for a given wireless channel, there will always be phase noise resulting from imperfect phase compensation at the receiving and transmitting RIS system. In this regard, there have been several works that collectively categorize phase noise to follow a Gaussian or a Generalized Uniform distribution \cite{Trigui_phasenoise}. It is important to note that these two distributions are the most versatile representations of phase as they cater to several acceptable assumptions. 
	
	It should be mentioned that the combined effect of fading and phase noise has been extensively studied in the context of RIS-RF \cite{Kudathanthirige2020, Qian_RIS_phase,LiDong_RIS_phase, wang_RIS_phase, Peng_RIS_phase,waqar2021performance}.  In \cite{Kudathanthirige2020}, \cite{Qian_RIS_phase}, \cite{LiDong_RIS_phase}, \cite{wang_RIS_phase}, authors have applied the central limit theorem (CLT) to derive an achievable diversity order, maximize SNR and quantifying capacity degradation of the RIS system. Although in\cite{Badiu_RIS_phase} , more accurate methods than CLT are used, authors approximated an  arbitrary fading model with Nakagami-$m$ distributed to develop performance analysis of the RIS-assisted transmission with phase noise. In the reference \cite{Peng_RIS_phase}, the diversity order of the system is derived under quantized, discrete values of phase shifts. The authors derived approximate performance bounds for the Rician fading model with phase errors in\cite{waqar2021performance}. In \cite{chapala2022phasenoise}, an exact performance on the RIS-RF for vehicular communications over $\kappa$-$\mu$ and double generalized Gamma fading has been presented.  To the best of authors' knowledge, there has been no previous work established on carrying out a performance analysis of RIS-aided THz systems under the combined  effects of channel fading, TPE,  and phase noise. 
	
	With the above motivation, in this paper, we analyze the performance of RIS-THz system under the combined effect of $\alpha$-$\mu$ fading channel, TPE, and statistical phase noise  in exact closed form expressions. We develop analytical expressions for the PDF and CDF of the resultant signal-to-noise ratio (SNR) of the RIS-THz link using multi-variate Fox's H-function by deriving the statistical results of the double $\alpha$-$\mu$ fading combined with THz pointing errors and phase noise at individual RIS element using single-variate Fox's H-function. Using the derived statistical results, we analyze the exact and asymptotic expressions for the outage probability and average bit-error-rate (BER) of the considered system. The analytical results show that the  diversity order of the system is independent of phase noise, depends on the channel fading parameters $\alpha$, $\mu$, and depends on the $\beta$ parameter of the TPE.



	\section{System Model}	
	We consider a single-antenna transceiver system assisted by an RIS with $N$ reflecting elements resulting into the received signal $y$ at the destination as
	\begin{eqnarray}\label{model_1}
		y = \sqrt[]{P_{t}} x \sum_{i=1}^{N} H_{l,i} \lvert h_{i}^{}\rvert  G_{l,i}^{}  \lvert g_{i}^{}\rvert  h_{p,i} e^{\J \theta_{i}} + v
	\end{eqnarray}
where $P_{t}$ is the transmit power, $x$ is the unit power information bearing signal, $v$ is the additive noise, and  $\lvert h_{i}^{}\rvert$  and  $\lvert g_{i}^{}\rvert$  are channel fading coefficients between the source to the $i$-th RIS element and between the $i$-th RIS element to the destination, respectively. Further,  $ H_{l,i}^{}$  and  $ G_{l,i}^{}$ denotes the channel gain of the two links. We denote the coefficient of pointer error as $h_{p,i}$ and the residual phase error at each RIS as $\theta_{i}$.

	We consider the channels from source to RIS and RIS to destination $\lvert h_{i}^{}\rvert$  and  $\lvert g_{i}^{}\rvert$ respectively, as independent but non-identically distributed (i.ni.d) $\alpha$-$\mu$ random variables. 	The $\alpha$-$\mu$ distribution is general, flexible, and has easy mathematical tractability. More importantly, experimental results of channel envelope at $142$ \mbox{GHz} shows excellent fit to the  $\alpha$-$\mu$ model. The PDF of $\lvert h_{i}^{}\rvert$ is given as \cite{Yacoub_alpha_mu} 
	\begin{eqnarray}
		\label{eq:h_i_pdf}
		f_{\lvert h_{i}^{}\rvert}(x) &= \frac{\alpha_{i,1}\mu_{i,1}^{\mu_{i,1}} x^{\alpha_{i,1}\mu_{i,1}-1}}{\Omega_{i,1}^{\alpha_{i,1}\mu_{i,1}}\times \Gamma(\mu_{i,1})} \times e^{-\mu_{i,1}\dfrac{ x^{\alpha_{i,1}}}{\Omega_{i,1}^{\alpha_{i,1}}}} 
	\end{eqnarray}
	where  $\Omega_{i,1}$ is the $\alpha_{i,1}$-root mean value of the fading channel envelope for the channel of the first link and the $i$-th RIS element with $\alpha_{i,1}$ and $\mu_{i,1}$ are non-linearity and multi-path clustering parameters.

For the TPE, we use a recently proposed statistical model in \cite{Dabiri_Pointing_Error_2022} with PDF as
	\begin{eqnarray}\label{eq:pe_pdf}
		f_{h_{p,i}}\left(x\right)=C_{i} \frac{x^{1 / \beta_{i,1}-1}}{G_{i,0}^{1 / \beta_{i,1}}} \sum_{k_{i}=0}^{K_{i}} \frac{\Delta_{i,k_{i}}\left(-\ln \left(\frac{x}{G_{i,0}}\right)\right)^{k_{i}+1}}{\Gamma(k_{i}+2) \beta_{i,1}^{k_{i}+2}}
	\end{eqnarray}
	for $0\leq x \leq G_{i,0}$, where
	$$
	\left\{\begin{array}{l}
		C_{i}=\prod_{j=1}^{4} \sqrt{\beta_{i,1} / \beta_{i,j}}, \\
		\Delta_{i,k_{i}}=\frac{1}{k_{i}} \sum_{j=1}^{k_{i}} j \gamma_{i,j} \Delta_{i,k_{i}+1-j} \text { for } k_{i}=1, \ldots, K_{i}, \\
		\Delta_{i,0}=1, \quad \gamma_{i,k_{i}}=\sum_{j=1}^{4} \frac{\left(1-\beta_{i,1} / \beta_{i,j}\right)^{k_{i}}}{2 k_{i}} .
	\end{array}\right.
	$$
	The TPE parameters used in \eqref{eq:pe_pdf} are defined comprehensively in \cite{Dabiri_Pointing_Error_2022}. As such, $\beta$ signifies the strength of pointing errors: more is the $\beta$ more is the pointing errors.
	
	To model the phase noise, we use the popular generalized uniform distribution i.e., $\theta_{i}\sim \mathcal{U}\left(-q\pi,q\pi\right)$ in our analysis, where $q$ denotes the quantization  level \cite{Trigui_phasenoise}.
	
	\section{Performance Analysis}
	In this section, we derive important statistical  results of the RIS-THz system to facilitate the outage probability and average BER  performances of the considered RIS-THz system under the combined effect of $\alpha$-$\mu$ channel fading, TPE,  and phase noise.  First, we derive the PDF and CDF of the double $\alpha$-$\mu$ fading combined with TPE and phase noise for a single RIS element captured through the random variable  $Z_{i}=\lvert h_{i}^{}\rvert \lvert g_{i}^{}\rvert h_{p,i} e^{\J \theta_{i}}$. Next, we analyze the random variable  $Z=\sum_{i=1}^{N}Z_{i}$ to develop statistical results considering the accumulating propagation effect from all RIS elements.

	In the following Lemma, we present the PDF of the combined effect of double $\alpha$-$\mu$ fading,  TPE,  and phase noise.
	\begin{my_lemma}
		The PDF of $Z_i$ for the single-element RIS propagation environment is given by
			\begin{eqnarray}\label{eq: Z_i_pdf_final}
		&f_{Z_{i}}(x)=\frac{x^{-1}}{\Gamma{(\mu_{i,1})} \Gamma{(\mu_{i,2})}} C_{i} \sum_{k_{i}=0}^{K_{i}} \frac{\Delta_{i,k_{i}}}{ \beta_{i,1}^{k_{i}+2}}  H_{k_{i}+3,k_{i}+5}^{k_{i}+4,0}  \nonumber \\ \! &\left[\!\psi_{i}x \middle\vert \!\! \begin{array}{c} \{(1+1 / \beta_{i,1},1)\}_{1}^{k_{i}+2},(1,q) \\ (\mu_{i,1},\frac{1}{\alpha_{i,1}}), (\mu_{i,2},\frac{1}{\alpha_{i,2}}),\{(1 / \beta_{i,1},1)\}_{1}^{k_{i}+2},(0,q) \end{array}\!\!\!\!\right]
		\end{eqnarray} 
	\end{my_lemma}
\begin{IEEEproof}
See Appendix A.
	\end{IEEEproof}
The result present in Lemma 1 is exact using a single-variate Fox's H-function, which can be efficiently computed using the recently introduced function FoxH in Mathematica. Note that the PDF of a direct link combined with a single $\alpha$-$\mu$ fading with TPE was presented in an  approximate form using Hypergeomteric function  \cite{Dabiri_Pointing_Error_2022}.

In the following Theorem, we  derive the PDF of the sum of the random variable  $Z=\sum_{i}^NZ_i$, where $Z_i$ is distributed according to \eqref{eq: Z_i_pdf_final}.  Considering the single-variate representation of $f_{Z_i}(x)$, the PDF $f_Z(x)$ requires $N$-variate  Fox's H-function:
	\begin{my_theorem}\label{th:ris_pdf_phase}
		The PDF of  $Z$ for the accumulated effect from all RIS elements is given as
		\begin{eqnarray}\label{eq:Z_pdf_final}
			&f_{Z_{\rm }}(x) =  \prod_{i=1}^{N}  \frac{x^{-1}}{\Gamma{(\mu_{i,1})} \Gamma{(\mu_{i,2})}} C_{i} \sum_{k_{i}=0}^{K_{i}} \frac{\Delta_{i,k_{i}}}{ \beta_{i,1}^{k_{i}+2}}
			\nonumber \\&H_{0,1:k_{1}+4,k_{1}+5;\cdots;k_{N}+4,k_{N}+5}^{0,0:k_{1}+4,1;\cdots;k_{N}+4,1} \nonumber \\ &\bigg[\begin{array}{c} \{\psi_{i}x\}_{1}^{N} \end{array} \big\vert \begin{array}{c}-:\{(1,1),U_{i},(1,q)\}_{i=1}^{N}\\ (1:1,\cdots,1):\{V_i,(0,q)\}_{i=1}^{N} \end{array}\bigg] 	
		\end{eqnarray}
				where $U_{i}=\{(1+1 / \beta_{i,1},1)\}_{1}^{k_{i}+2}$ and $V_i =  (\mu_{i,1},\frac{1}{\alpha_{i,1}}) , (\mu_{i,2},\frac{1}{\alpha_{i,2}}), \{(1 / \beta_{i,1},1)\}_{1}^{k_{i}+2}$.
		\end{my_theorem}
	
	\begin{IEEEproof}
		See appendix B.
	\end{IEEEproof}
The multi-variate Fox's H-function can be efficiently computed using computational software and provides an elegant asymptotic analysis facilitating tuning of system parameters for the deployment of RIS-THz systems.    The use of multi-variate Fox's H-function approach for other RIS-assisted wireless systems has been employed in recent works \cite{du2020_thz, chapala2021unified, chapala2021THz}.	

In what follows, we  use the results of Theorem 1 to derive exact expressions for the outage probability and average BER of RIS-THz system. To do so, we required the PDF $f_{\gamma}(\gamma)$  of the resultant SNR $\gamma =\bar{\gamma}|Z|^2$ of the RIS-THz link, which  can be expressed using mathematical transformation as	
		$f_{\gamma_{\rm}^{}}(\gamma)=\frac{1}{2\sqrt{\bar{\gamma}_{\rm }\gamma}}f_{Z_{\rm }^{}}(\sqrt{\frac{\gamma}{\bar{\gamma}_{\rm }}}) $, where $\bar{\gamma}$ is the average SNR.
		
Mathematically, the outage probability is the probability of SNR falling below a SNR threshold value $\gamma_{\rm th}$ i.e.,  $ P_{\rm out}=Pr(\gamma \le \gamma_{\rm th})$.
	\begin{my_lemma}
Analytical expressions for exact and asymptotic expression for the outage probability of RIS-THz system is given by
			\begin{eqnarray}\label{eq:SNR_cdf_final}
		&P_{\rm out} =  \prod_{i=1}^{N}  \frac{1}{\Gamma{(\mu_{i,1})} \Gamma{(\mu_{i,2})}} C_{i} \sum_{k_{i}=0}^{K_{i}} \frac{\Delta_{i,k_{i}}}{ \beta_{i,1}^{k_{i}+2}}
		\nonumber \\&H_{0,1:k_{1}+4,k_{1}+5;\cdots;k_{N}+4,k_{N}+5}^{0,0:k_{1}+4,1;\cdots;k_{N}+4,1} \nonumber \\ &\bigg[\begin{array}{c} \{\psi_{i}\sqrt{\frac{\gamma_{\rm th}}{\bar{\gamma_{\rm }}_{\rm }}}\}_{1}^{N} \end{array} \big\vert \begin{array}{c}-:\{(1,1),U_{i},(1,q)\}_{i=1}^{N}\\ (0:1,\cdots,1):\{V_i,(0,q)\}_{i=1}^{N} \end{array}\bigg] 	
		\end{eqnarray}
		where $U_{i}=\{(1+1 / \beta_{i,1},1)\}_{1}^{k_{i}+2}$ and $V_i =  (\mu_{i,1},\frac{1}{\alpha_{i,1}}) , (\mu_{i,2},\frac{1}{\alpha_{i,2}}), \{(1 / \beta_{i,1},1)\}_{1}^{k_{i}+2}$.

		\begin{eqnarray}\label{eq:outage_asymp}
		&P_{{\rm out}}^{\infty} = \prod_{i=1}^{N}  \frac{1}{\Gamma{(\mu_{i,1})} \Gamma{(\mu_{i,2})}} C_{i} \sum_{k_{i}=0}^{K_{i}} \nonumber\\ &\frac{\Delta_{i,k_{i}}}{ \beta_{i,1}^{k_{i}+2}} \frac{1}{\Gamma(1+\frac{p_{1}}{2}+\cdots+\frac{p_{N}}{2})}   \prod_{i=1}^{N} \dfrac{(\Gamma({1/\beta_{i,1}+p_{i}}))^{(k_{i}+2)}}{(\Gamma({1+/\beta_{i,1}+p_{i}}))^{(k_{i}+2)}} \nonumber \\ &\dfrac{\Gamma({\mu_{1,i}-\frac{p_{i}}{\alpha_{1,i}})}  \Gamma({\mu_{2,i}-\frac{p_{i}}{\alpha_{2,i}}})}{\Gamma(1+qp_{i})\Gamma(1-qp_{i})}  \Gamma (p_{i})(\psi_{i} \sqrt{\frac{\gamma_{\rm th}}{\bar{\gamma}_{\rm }}}^{})^{p_{i}}
		\end{eqnarray}
		\end{my_lemma}
	\begin{IEEEproof}	
	We use standard procedure to integrate the multi-variate Fox's H-function in \eqref{th:ris_pdf_phase} to get the CDF $F_Z(x)=\int_0^x f_Z(x)$. Applying the transformation of random variable, we get  $P_{\rm out}=F_{\gamma_{\rm }^{}}(\gamma_{\rm th})=F_{Z_{\rm }^{}}(\sqrt{\frac{\gamma_{\rm th}}{\bar{\gamma}_{\rm }}})$ in \eqref{eq:SNR_cdf_final}.

We compute the residue of \eqref{eq:SNR_cdf_final} at a dominate pole $p_{i} = \min\{\alpha_{i,1}\mu_{i,1},\alpha_{i,2}\mu_{i,2},\{1 / \beta_{i,1}\}_{1}^{k_{i}+2}\}$ \cite[eq. (30)]{Rahama2018}	to get the outage probability at a high SNR in \eqref{eq:outage_asymp}. 
	
	\end{IEEEproof}
The asymptotic expression in \eqref{eq:outage_asymp} allows to develop the diversity order of the system as $G_{\rm out}= \sum_{i}^{N}\frac{p_{i}}{2}= \sum_{i}^{N} \min\{\alpha_{i,1}\mu_{i,1},\alpha_{i,2}\mu_{i,2},\{1 / \beta_{i,1}\}_{1}^{k_{i}+2}\} $. The diversity order expression reveals some interesting facts: the diversity order is independent of phase noise, and depends on the beam-width parameter $\beta$ of the TPE together with channel parameters $\alpha$ and $\mu$. Thus, the diversity order provides various design criteria depending on the fading and system parameters.
	
Finally, we analyze the average BER performance defined as  \cite{Ansari2011_ber}
	\begin{equation}\label{eq:gen_ber}
		\bar{P}_{e} =  \frac{q^{p}}{2\Gamma(p)} \int_{0}^{\infty} \gamma^{p-1} e^{-q\gamma} F_{\gamma_{\rm }} (\gamma) d\gamma
	\end{equation}
	where $p$ and $q$ are modulation specific parameters.
\begin{my_lemma}
	Analytical expressions for exact and asymptotic expression for the average BER of the RIS-THz system is given by
	\begin{eqnarray}\label{eq:ber_final}
	&\bar{P}_{e} =  \frac{1}{2\Gamma(p)} \prod_{i=1}^{N}  \frac{1}{\Gamma{(\mu_{i,1})} \Gamma{(\mu_{i,2})}} C_{i} \sum_{k_{i}=0}^{K_{i}} \frac{\Delta_{i,k_{i}}}{ \beta_{i,1}^{k_{i}+2}} \nonumber \\&H_{1,1:k_{1}+4,k_{1}+5;\cdots;k_{N}+4,k_{N}+5}^{0,1:k_{1}+4,1;\cdots;k_{N}+4,1} \nonumber \\ &\hspace{-6mm}\left[\!\!\!\begin{array}{c} \{\frac{\psi_{i}}{\sqrt{q\bar{\gamma}_{\rm }}}\}_{1}^{N} \end{array}\!\! \middle\vert \begin{array}{c}(1-p:\frac{1}{2},\cdots,\frac{1}{2}):\{(1,1),U_{i},(1,q)\}_{i=1}^{N}\\ (0:1,\cdots,1):\{V_i,(0,q)\}_{i=1}^{N} \end{array}\!\!\right] 
	\end{eqnarray}
	where $U_{i}=\{(1+1 / \beta_{i,1},1)\}_{1}^{k_{i}+2}$ and $V_i =  (\mu_{i,1},\frac{1}{\alpha_{i,1}}) , (\mu_{i,2},\frac{1}{\alpha_{i,2}}), \{(1 / \beta_{i,1},1)\}_{1}^{k_{i}+2}$.
	
		\begin{eqnarray}\label{eq:ber_asymp}
	&\bar{P}_{e}^{\infty} = \frac{1}{2\Gamma(p)}\prod_{i=1}^{N}  \frac{1}{\Gamma{(\mu_{i,1})} \Gamma{(\mu_{i,2})}} C_{i} \sum_{k_{i}=0}^{K_{i}} \frac{\Delta_{i,k_{i}}}{ \beta_{i,1}^{k_{i}+2}} \nonumber\\ &\frac{\Gamma(p+\frac{p_{1}}{2}+\cdots+\frac{p_{N}}{2})}{\Gamma(1+\frac{p_{1}}{2}+\cdots+\frac{p_{N}}{2})}   \prod_{i=1}^{N} \dfrac{(\Gamma({1/\beta_{i,1}+p_{i}}))^{(k_{i}+2)}}{(\Gamma({1+/\beta_{i,1}+p_{i}}))^{(k_{i}+2)}} \nonumber \\ &\dfrac{\Gamma({\mu_{1,i}-\frac{p_{i}}{\alpha_{1,i}})}  \Gamma({\mu_{2,i}-\frac{p_{i}}{\alpha_{2,i}}})\Gamma (p_{i})}{\Gamma(1+qp_{i})\Gamma(1-qp_{i})}  (\psi_{i} \sqrt{\frac{1}{q\bar{\gamma}_{\rm }}}^{})^{p_{i}}
	\end{eqnarray}
	where $p_{i} = \min\{\alpha_{i,1}\mu_{i,1},\alpha_{i,2}\mu_{i,2},\{1 / \beta_{i,1}\}_{1}^{k_{i}+2}\}$.
\end{my_lemma}	
\begin{IEEEproof}
	We use \eqref{eq:SNR_cdf_final} in \eqref{eq:gen_ber} and then expand the definition of multi-variate Fox's H-function to solve the inner integral as $I=\int_{0}^{\infty} \gamma^{p-1} e^{-q\gamma} \gamma^{\sum_{i=1}^{N}\frac{s_{i}}{2}} d\gamma=(\frac{1}{q})^{p+\sum_{i=1}^{N}\frac{s_{i}}{2}} \Gamma(p+\sum_{i=1}^{N}\frac{s_{i}}{2})$, and apply definition of multi-variate Fox's H-function \cite{M-Foxh} to get  \eqref{eq:ber_final}.
	
Applying the similar procedure for the asymptotic analysis of  outage probability (as given in the proof of Lemma 2), we get  the average BER at a high SNR in \eqref{eq:ber_asymp}.
\end{IEEEproof}
The diversity order of the average BER has similar functional representation as that of the diversity order for the outage probability.

It should be emphasized that multi-variate Fox's H-function provides a nice asymptotic expansion in terms of simpler algebraic and Gamma functions. 
 	\section{Simulation and Numerical Results}\label{sec:sim_num_res}
	In this section, we use numerical analysis and Monte Carlo simulations  to demonstrate the effect of phase noise, TPE, and channel fading on the performance of RIS-assisted THz transmission. We also  validate the derived analytical expressions using simulations. For numerical computation of the derived analytical expressions, we use the  Python code implementation of multi-variate Fox's H-function \cite{Alhennawi2016}. 
	
	We consider  THz carrier frequency at  $300$\mbox{GHz}, transmit antenna gain $G_{T}=40$\mbox{dBi}, and receive antenna gain $G_{R}=40$\mbox{dBi}, and molecular absorption coefficient $k=3.18 \times 10^{-4}$ per meter. We assume the distance from the source to RIS at $d_1 = 20$\mbox{m} and the RIS to destination at $d_2=50$\mbox{m}.   We consider $\alpha_{i,1}=\alpha_{i,2}=\alpha=2$, $\mu_{i,1}=\mu_{i,2}=\mu=1$ $\forall i$  and $\alpha_{i,1}=\alpha_{i,2}=\alpha=2$, $\mu_{i,1}=\mu_{i,2}=\mu=4$ $\forall i$ fading parameters in our simulations. We use pointing error parameters as described in \cite{Dabiri_Pointing_Error_2022} to get $\beta_{q_{i,w}}=\frac{1}{12}$. A noise floor of $-74$\mbox{dBm} is considered over a $10$\mbox{GHz} channel bandwidth.
	
	\begin{figure*}
		\centering
		\subfigure[Without pointing errors.]{\includegraphics[scale=0.4]{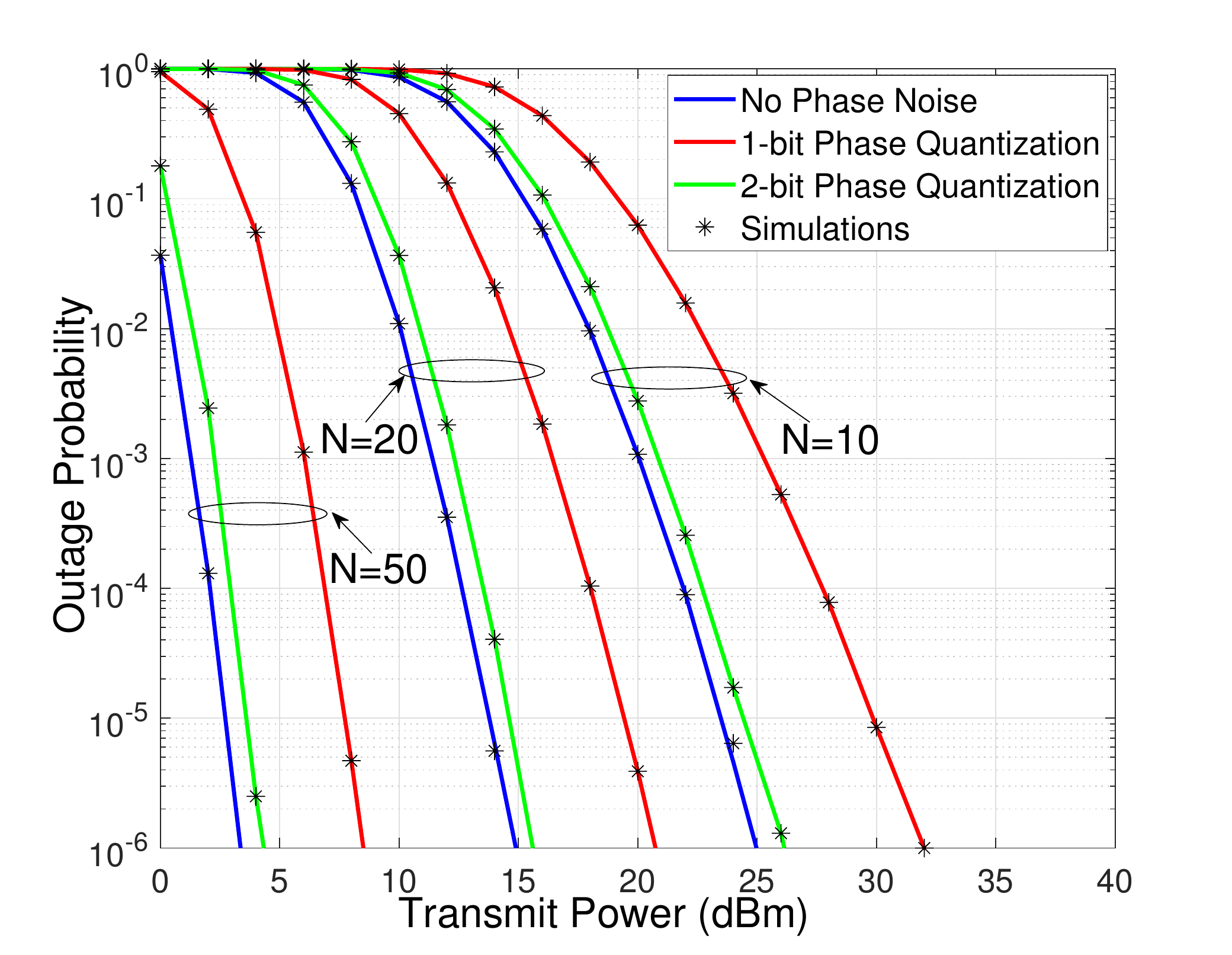}} 
		\subfigure[With pointing errors.]{\includegraphics[scale=0.4]{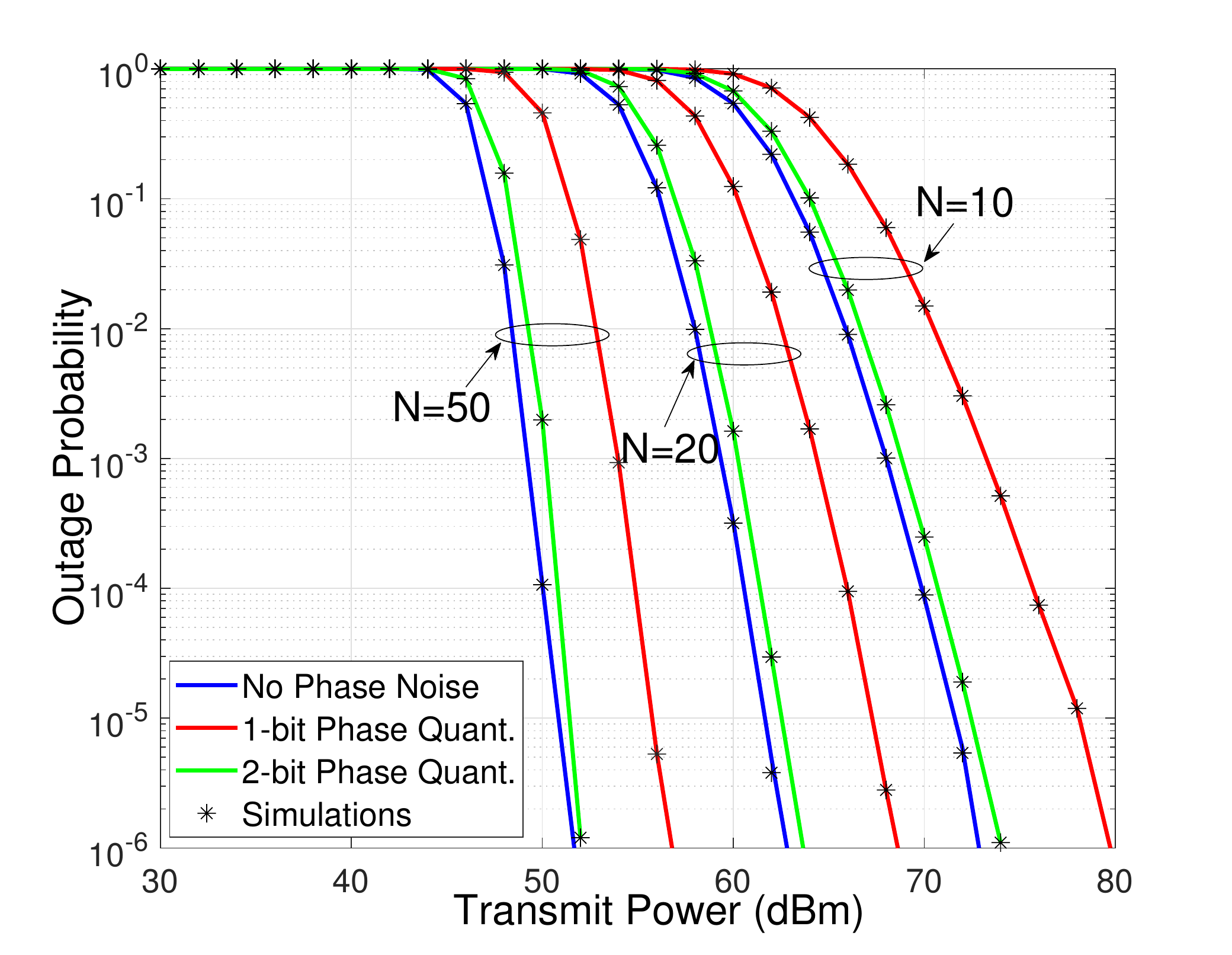}} 
		\caption{Outage performance of  the RIS-THz system  with $\alpha=2$, $\mu=1$ and different phase noise levels.}
		\label{fig:outage_ris}
	\end{figure*}
	
	Fig.~\ref{fig:outage_ris} plots the outage performance of the considered THz system for $\alpha=2$, $\mu=1$ combined with the TPE and phase noise for different RIS elements $N$. We also plot the outage probability  without TPE in Fig.~\ref{fig:outage_ris}(a) to illustrate the effect of phase noise. We consider $1$ bit and $2$ bits quantization levels to represent the phase noise and compare it with that of perfect phase compensation scenario. With $1$-bit quantization, we can observe a $4\mbox{dBm}$ loss in transmit power to achieve a desired outage performance of $10^{-3}$ with $N=10$. However, the performance improves with an increase in the level of phase compensation achieving the outage probability  similar to that of the perfect phase compensation scenario.However, a large RIS size can be used to compensate the loss in performance due to imperfect phase estimation. In Fig.~\ref{fig:outage_ris}(b), with $N=20$ and $1$-bit quantization, a desired outage performance of $10^{-3}$ can be achieved at $P_{t}=64\mbox{dBm}$ when compared to $P_{t}=69\mbox{dBm}$ with $N=10$ and $2$-bit quantization. Further, the gain in transmit power is significant for $N=50$ and $1$-bit quantizer. Thus, the proposed analysis gives a design criteria to appropriately choose the phase noise quantization bits and the number of RIS elements for a given desired performance metric.

\begin{figure*}
		\centering
		\subfigure[Without pointing errors.]{\includegraphics[scale=0.4]{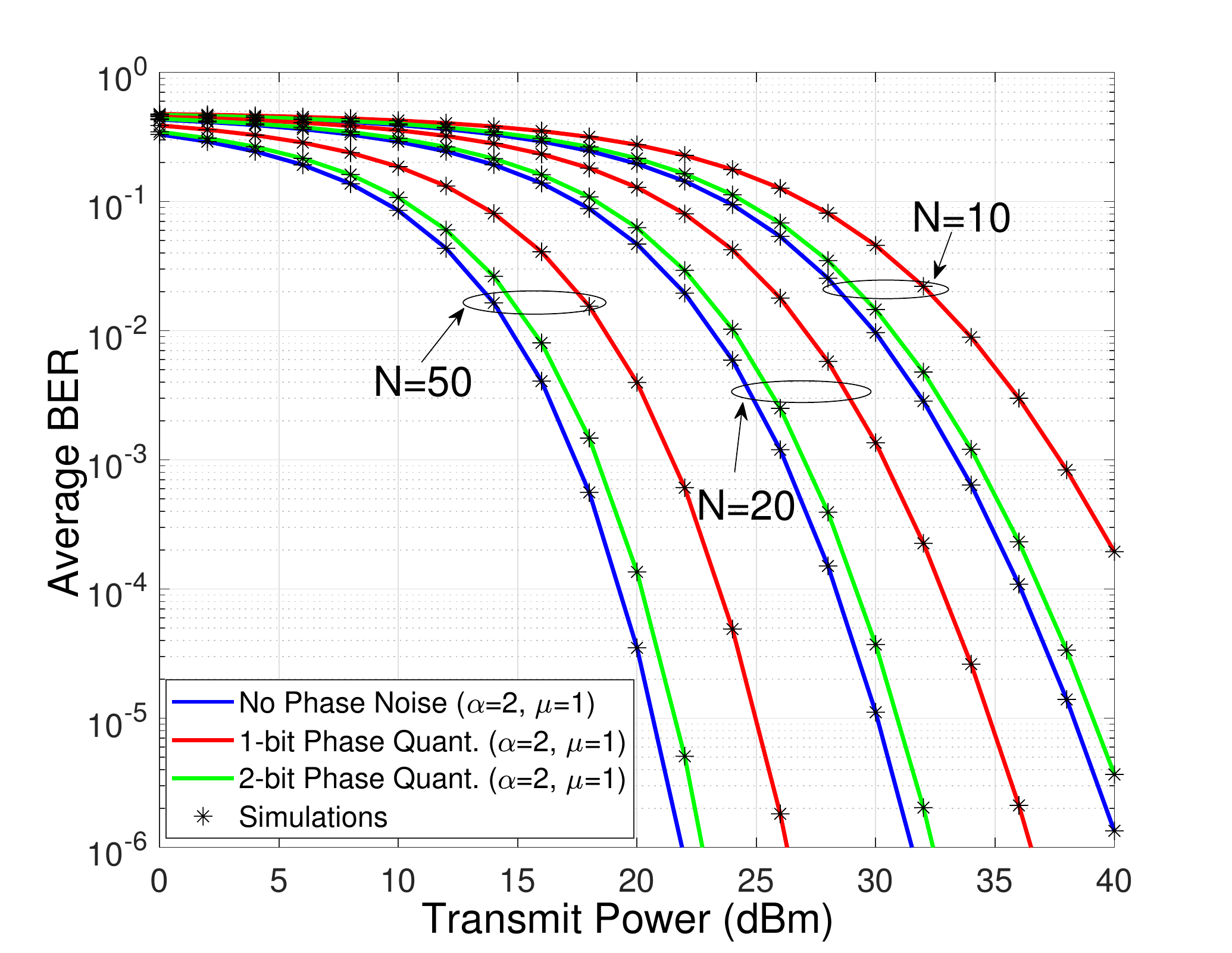}} 
		\subfigure[With pointing errors.]{\includegraphics[scale=0.4]{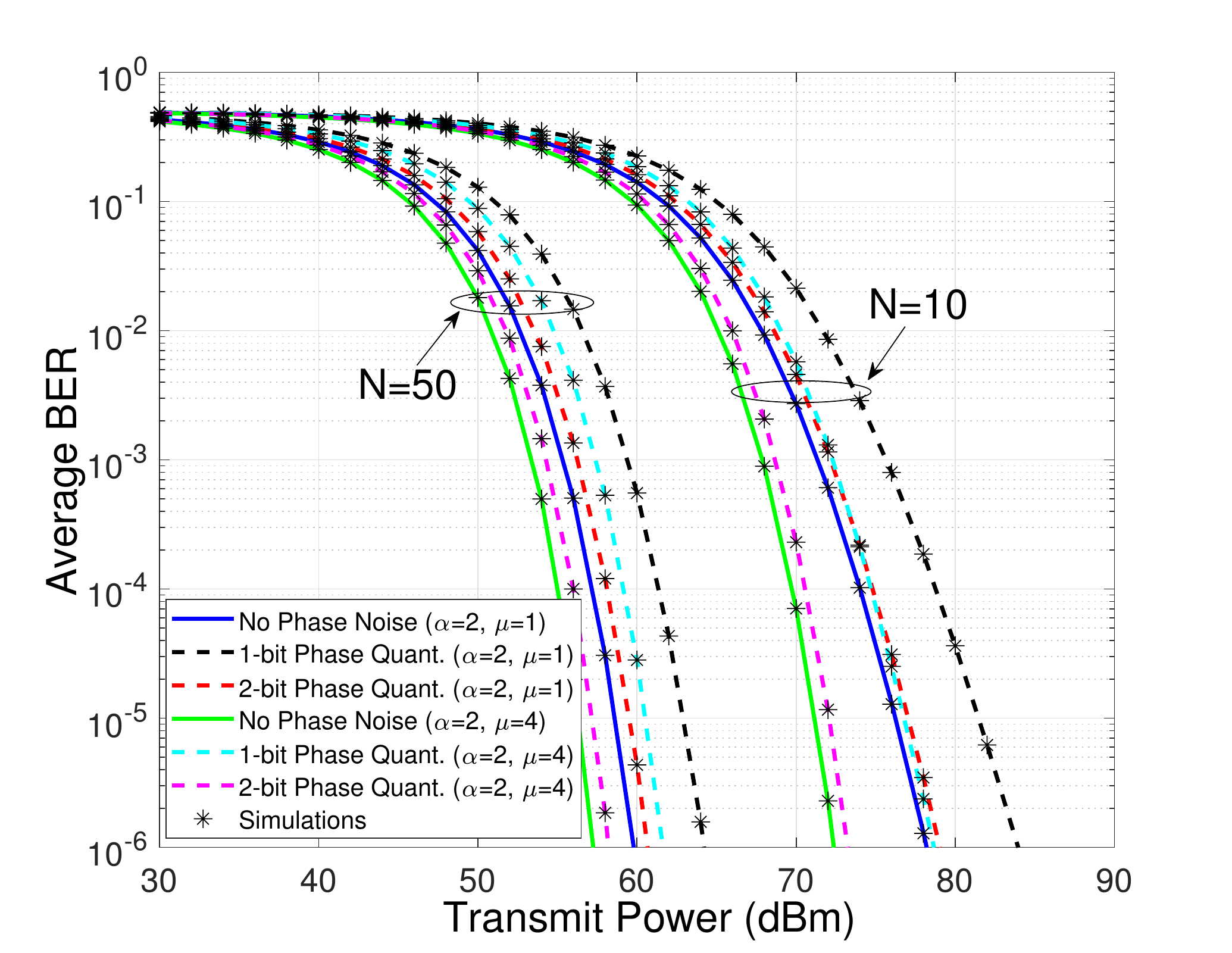}} 
		\caption{Average BER of the RIS-THz system for different fading parameters and phase noise levels.}
		\label{fig:ber_ris_mob}
	\end{figure*}

	Fig.~\ref{fig:ber_ris_mob} shows the average BER performance for different fading parameters, RIS size ($N$) and  different phase noise quantization bits  without TPE Fig.~\ref{fig:ber_ris_mob} (a) and with the TPE Fig.~\ref{fig:ber_ris_mob}(b). We  consider different fading parameters to depict the dependence of the diversity order depends on system parameters i.e., for $\alpha=2$, $\mu=1$, BER diversity order $N$ and for $\alpha=2$, $\mu=4$, BER diversity order $4N$. Further, it can be seen that the BER performance improves with an increase in the multi-path clustering parameter $\mu$. Similar to the outage performance, BER improves with an increase in quantization level for a given $N$. Also, the desired performance can be achieved with larger $N$  even with a lower quantization level or smaller $N$ with a higher quantization level. 
	
	In all above plots, simulation results match with  to numerical analysis,  validating the analytical results presented in this paper.
	
	\section{Conclusion}
	 In this paper, we analyzed the performance of RIS-THz transmission under the combined effect of channel fading, TPE, and statistical phase noise due to imperfect phase compensation at each RIS element. We  employed a novel and scalable approach to derive the statistical PDF and CDF of effective RIS channels under  the combined effects of three channel impairments modeled statistically. We  presented exact closed form expressions for outage and average BER of the considered system in terms of multi-variate Fox's H-function. We also developed asymptotic analysis to reveal that the
	  the diversity order is independent of phase noise, and depends on the beam-width parameter $\beta$ of the TPE together with channel parameters $\alpha$ and $\mu$. 	 
	  The proposed analysis provides a design criteria to appropriately choose RIS size ($N$) and phase noise compensator to achieve a desired system performance under fading channel. A near-optimal performance under the phase noise can be achieved by employing larger $N$  with a lower quantization level or smaller $N$ with a higher quantization level for phase noise compensator at the RIS. 
	  
	  The proposed analysis can be augmented to analyze the performance of more advanced THz wireless networks.
	\section*{Appendix A} 
Expressing the exponential function of 	\eqref{eq:h_i_pdf} using single-variate Fox's H-function in $f_{Z_{hg,i}}(x)=\int_{0}^{\infty} \frac{1}{u} f_{\lvert h_{i}^{}\rvert}(\frac{x}{u}) f_{\lvert g_{i}^{}\rvert}(u) du$ and apply the identity  \cite[Th 2.3]{M-Foxh}  to develop the PDF of double $\alpha$-$\mu$ fading $Z_{hg,i} = \lvert h_{i}^{}\rvert \lvert g_{i}^{}\rvert$ as  \cite{Badarneh2020_double}
	\begin{eqnarray}\label{eq:hg_i_pdf}
	&f_{Z_{hg,i}}(x) = \frac{x^{-1} }{\Gamma{(\mu_{i,1})} \Gamma{(\mu_{i,2})}} \nonumber \\ &H_{0,2}^{2,0}  \left[\dfrac{x}{\mathcal{A}_{i,1}\mathcal{A}_{i,2}} \middle\vert \begin{array}{c} - \\ (\mu_{i,1},\frac{1}{\alpha_{i,1}}) , (\mu_{i,2},\frac{1}{\alpha_{i,2}}) \end{array}\right]
	\end{eqnarray}
	where $\mathcal{A}_{i,1}=\frac{\Omega_{i,1}}{\mu_{i,1}^{\frac{1}{\alpha_{i,1}}}}$, $\mathcal{A}_{i,2}=\frac{\Omega_{i,2}}{\mu_{i,2}^{\frac{1}{\alpha_{i,2}}}}$.
	
Next, we combine the TPE with the double  $\alpha$-$\mu$ fading $Z_{hg,i}$ using the  product of two random variables formula to derive the PDF of $Z_{phg,i}=h_{p,i}Z_{hg,i}$ as
	\begin{flalign}\label{eq:hgp_i_pdf_1}
	&f_{Z_{phg,i}}(x)=\int_{\frac{x}{G_{i,0}}}^{\infty} \frac{1}{u} f_{\lvert h_{p,i}^{}\rvert}(\frac{x}{u}) f_{\lvert Z_{hg,i}^{}\rvert}(u) du \nonumber \\
	&= \frac{1}{\Gamma{(\mu_{i,1})} \Gamma{(\mu_{i,2})}} C_{i} \frac{x^{1 / \beta_{i,1}-1}}{G_{i,0}^{1 / \beta_{i,1}}} \sum_{k_{i}=0}^{K_{i}} \frac{\Delta_{i,k_{i}}}{\Gamma(k_{i}+2) \beta_{i,1}^{k_{i}+2}} \nonumber \\ &\int_{\frac{x}{G_{i,0}}}^{\infty} \left(-\ln \left(\frac{x}{uG_{i,0}}\right)\right)^{k_{i}+1} u^{-1 / \beta_{i,1}} u^{-1} \nonumber\\&H_{0,2}^{2,0}  \left[\dfrac{u}{\mathcal{A}_{i,1}\mathcal{A}_{i,2}} \middle\vert \begin{array}{c} - \\ (\mu_{i,1},\frac{1}{\alpha_{i,1}}) , (\mu_{i,2},\frac{1}{\alpha_{i,2}}) \end{array}\right] du
	\end{flalign}
Expanding the  definition of Fox's H-function, solving the resultant inner integral with the  substitution $y=ln\left(\dfrac{uG_{i,0}}{x}\right)$ as	$I= (\frac{x}{G_{i,0}})^{-1 / \beta_{i,1}-s_{i}} (\frac{\Gamma(1 / \beta_{i,1}+s_{i})}{\Gamma(1+1 / \beta_{i,1}+s_{i})})^{k_{i}+2} \Gamma(k_{i}+2)$, we get
	\begin{eqnarray}\label{eq:hgp_i_pdf}
	&f_{Z_{phg,i}}(x)= \frac{x^{-1}}{\Gamma{(\mu_{i,1})} \Gamma{(\mu_{i,2})}} C_{i} \sum_{k_{i}=0}^{K_{i}} \frac{\Delta_{i,k_{i}}}{ \beta_{i,1}^{k_{i}+2}}  \nonumber \\&\hspace{-6mm}H_{k_{i}+2,k_{i}+4}^{k_{i}+4,0}\!\!  \left[\psi_{i}x \middle\vert \!\!\begin{array}{c} \{(1+1 / \beta_{i,1},1)\}_{1}^{k_{i}+2} \\ (\mu_{i,1},\frac{1}{\alpha_{i,1}}) , (\mu_{i,2},\frac{1}{\alpha_{i,2}}), \{(1 / \beta_{i,1},1)\}_{1}^{k_{i}+2} \end{array}\!\!\!\right]
	\end{eqnarray}
	where $\psi_{i}=\dfrac{1}{G_{i,0}\mathcal{A}_{i,1}\mathcal{A}_{i,2}}$.
	
	Finally, we combine the result in \eqref{eq:hgp_i_pdf}  with phase noise to get PDF of $Z_i = Z_{phg,i}\cdot e^{\J\theta_i}$. We apply the approach of \cite{chapala2022phasenoise} to develop the PDF of $Z_i$ in terms of a univariate Fox H-function. We compute PDF $Z_i$ as $Z_{phg,i} \cdot e^{\J\theta_i}$ by making use of conditional expectation of random variables \cite{chapala2022phasenoise}:
	\begin{eqnarray}\label{eq: Conditional Expectation z_i}
	f_{Z_{i}^{}}(x) = \Aver{f_{Z_{i}^{}}(x/\theta_{i})} = 
	= \int_{-q\pi}^{q\pi} \dfrac{1}{e^{\J\theta_i}} f_{Z_{phg,i}}(\frac{x}{e^{\J \theta_{i}}})f_{\theta_{i}}(\theta) d\theta 
	\end{eqnarray}
	For a given constant $\theta_{i}$, the probability distribution of $Z_i$ simplifies into:
	\begin{eqnarray}\label{eq: Z_i_pdf_1}
	&f_{Z_{i}^{}}(x/\theta_{i}) = e^{-\J\theta_i} f_{Z_{phg,i}}(\frac{x}{e^{\J \theta_{i}}}) = \frac{x^{-1}}{\Gamma{(\mu_{i,1})} \Gamma{(\mu_{i,2})}} C_{i} \nonumber \\&\sum_{k_{i}=0}^{K_{i}} \frac{\Delta_{i,k_{i}}}{ \beta_{i,1}^{k_{i}+2}}  H_{k_{i}+2,k_{i}+4}^{k_{i}+4,0}  \nonumber \\ &\left[\psi_{i}xe^{-\J \theta_{i}} \middle\vert \!\! \begin{array}{c}\{(1+1 / \beta_{i,1},1)\}_{1}^{k_{i}+2} \\ (\mu_{i,1},\frac{1}{\alpha_{i,1}}) , (\mu_{i,2},\frac{1}{\alpha_{i,2}}), \{(1 / \beta_{i,1},1)\}_{1}^{k_{i}+2}\end{array}\!\!\!\right]
	\end{eqnarray}
Thus, substituting \eqref{eq: Z_i_pdf_1} in \eqref{eq: Conditional Expectation z_i} and expanding the definition of Fox's H-function with $\Aver{e^{-\J \theta_{i} s_{i}}} = \frac{\sin{q\pi s_{i}}}{q\pi s_{i}} = \frac{1}{\Gamma(1-q s_{i}) \Gamma(q s_{i})} \frac{\Gamma(q s_{i})}{\Gamma(1+q s_{i})} = \frac{1}{\Gamma(1-q s_{i})\Gamma(1+q s_{i})}$, we get \eqref{eq: Z_i_pdf_final}, which completes the proof of Lemma 1.
	
	\section*{Appendix B}
	The PDF for $Z_{\rm}$ can derived  using the moment generating function (MGF) and inverse laplace transform in succession as $f_{Z_{\rm RIS}^{}}(z) = \mathcal{L}^{-1} \prod_{i=1}^{N} M_{Z_i^{}}(s)$, where
	\begin{eqnarray}\label{eq: mgf_Z_i_1}
		&M_{Z_i^{}}(s) =\int_{0}^{\infty}e^{-sx} f_{Z_{i}^{}}(x) dx \nonumber \\
		&=\frac{1}{\Gamma{(\mu_{i,1})} \Gamma{(\mu_{i,2})}} C_{i} \sum_{k_{i}=0}^{K_{i}} \frac{\Delta_{i,k_{i}}}{ \beta_{i,1}^{k_{i}+2}}\int_{0}^{\infty}x^{-1}e^{-sx} H_{k_{i}+3,k_{i}+5}^{k_{i}+4,0}  \nonumber \\&\hspace{-4mm}\left[\psi_{i}x \middle\vert \!\! \begin{array}{c} \{(1+1 / \beta_{i,1},1)\}_{1}^{k_{i}+2},(1,q) \\ (\mu_{i,1},\frac{1}{\alpha_{i,1}}) , (\mu_{i,2},\frac{1}{\alpha_{i,2}}), \{(\frac{1}{  \beta_{i,1}},1)\}_{1}^{k_{i}+2},(0,q) \end{array} \!\!\!\right]\!\!dx
	\end{eqnarray}  
	On expanding the Fox's H-function to the contour integral and interchanging the order of integration, we obtain
	\begin{eqnarray}\label{eq: mgf_Z_i_2}
		&M_{Z_i^{}}(s) = \frac{1}{\Gamma{(\mu_{i,1})} \Gamma{(\mu_{i,2})}} C_{i} \sum_{k_{i}=0}^{K_{i}} \frac{\Delta_{i,k_{i}}}{ \beta_{i,1}^{k_{i}+2}} \frac{1}{2\pi \J} \int_{\mathcal{L}_{i}} \left(\psi_{i}\right)^{s_i} \nonumber \\ &\dfrac{(\Gamma({1/\beta_{i,1}+s_{i}}))^{(k_{i}+2)}}{(\Gamma({1+/\beta_{i,1}+s_{i}}))^{(k_{i}+2)}} \dfrac{\Gamma({\mu_{1,i}-\frac{s_i}{\alpha_{1,i}})}  \Gamma({\mu_{2,i}-\frac{s_i}{\alpha_{2,i}}})}{\Gamma(1+qs_i)\Gamma(1-qs_i)}\nonumber \\&\int_{0}^{\infty} e^{-sx} x^{s_{i}-1} dx	 ds_i   
	\end{eqnarray} 
	We solve the inner integral using the known identity \cite[3.381.4]{integrals} as $\int_{0}^{\infty} e^{-sx} x^{s_{i}-1} dx = s^{-s_i}\Gamma (s_i)$. The MGF of a sum of RVs is given as a product of their respective MGFs, from which the PDF can be extracted back
	\begin{eqnarray}\label{eq:Z_pdf_1}
		&f_{Z_{\rm RIS}^{}}(x) = \mathcal{L}^{-1} \prod_{i=1}^{N} M_{Z_i^{}}(s)\nonumber \\
		&=\prod_{i=1}^{N}  \frac{1}{\Gamma{(\mu_{i,1})} \Gamma{(\mu_{i,2})}} C_{i} \sum_{k_{i}=0}^{K_{i}} \frac{\Delta_{i,k_{i}}}{ \beta_{i,1}^{k_{i}+2}} \frac{1}{2\pi \J} \int_{\mathcal{L}_{i}} \left(\psi_{i}\right)^{s_i} \nonumber \\ &\dfrac{(\Gamma({1/\beta_{i,1}+s_{i}}))^{(k_{i}+2)}}{(\Gamma({1+/\beta_{i,1}+s_{i}}))^{(k_{i}+2)}} \dfrac{\Gamma({\mu_{1,i}-\frac{s_i}{\alpha_{1,i}})}  \Gamma({\mu_{2,i}-\frac{s_i}{\alpha_{2,i}}})}{\Gamma(1+qs_i)\Gamma(1-qs_i)}\nonumber \\&\Gamma (s_i)\left(\frac{1}{2\pi \J} \int_{\mathcal{L}} s^{-\sum_{i}^{N}s_{i}} e^{sx} ds\right) ds_{i}
	\end{eqnarray}
	Applying the identity \cite[8.315.1]{integrals}, the inner integral can be solved as
	\begin{equation}\label{eq:Z_pdf_int_1}
		\frac{1}{2\pi \J} \int_{\mathcal{L}} s^{-\sum_{i}^{N}s_{i}} e^{sx} ds = \Big(\frac{1}{x}\Big)^{-\sum_{i}^{N}s_{i}+1} \frac{1}{\Gamma(\sum_{i}^{N}s_{i})}
	\end{equation}
Substituting \eqref{eq:Z_pdf_int_1} in \eqref{eq:Z_pdf_1} and	making the use of the linear combination of $N$ variables of $s_i$ with  the multi-variate Fox's H-function definition to get the PDF of the RIS-THz channel in \eqref{eq:Z_pdf_final}.

	
	\bibliographystyle{IEEEtran}
	\bibliography{Multi_RISE_full}
	
\end{document}